\documentstyle[12pt,epsfig]{article}                                                    

\newlength{\dinwidth}                                                    
\newlength{\dinmargin}                                                    
\def\lapproxeq{\lower .7ex\hbox{$\;\stackrel{\textstyle                                                    
<}{\sim}\;$}}                                                    
\def\gapproxeq{\lower .7ex\hbox{$\;\stackrel{\textstyle                                                    
>}{\sim}\;$}}                                                    
\def\be{\begin{equation}}                                                    
\def\ee{\end{equation}}                                                    
\def\bea{\begin{eqnarray}}                                                    
\def\eea{\end{eqnarray}}

\begin{document}                                                    
\titlepage                                                    
\begin{flushright}                                                    
IPPP/08/76   \\
DCPT/08/152 \\                                                    
\today \\                                                    
\end{flushright}                                                    
                                                    
\vspace*{2cm}                                                    
                                                    
\begin{center}                                                    
{\Large \bf Rapidity gap survival probability and total cross sections\footnote{Based on a talk by A.D. Martin
at the CERN - DESY Workshop "HERA and the LHC",
26 - 30 May 2008,
CERN. } }                                                    
                                                    
\vspace*{1cm}                                                    
A.D. Martin$^a$ ,V.A. Khoze$^{a,b}$ and M.G. Ryskin$^{a,b}$ \\                                                    
                                                   
\vspace*{0.5cm}                                                    
$^a$ Institute for Particle Physics Phenomenology, University of Durham, Durham, DH1 3LE \\                                                   
$^b$ Petersburg Nuclear Physics Institute, Gatchina, St.~Petersburg, 188300, Russia            
\end{center}                                                    
                                                    
\vspace*{2cm}                                                    

\begin{abstract}
We discuss recent calculations of the survival probability of the large rapidity gaps in exclusive processes of the type $pp \to p+A+p$ at high energies. Absorptive or screening effects are important, and one consequence is that the total cross section at the LHC is predicted to be only about 90 mb.
\end{abstract}

\begin{figure} 
\begin{center}
\includegraphics[height=5cm]{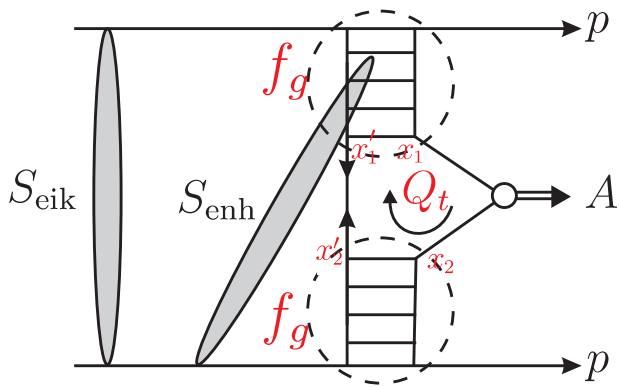}
\caption{The mechanism for the exclusive process $pp \to p+A+p$, with the eikonal and enhanced survival factors shown symbolically.} 
\label{fig:pAp}
\end{center}
\end{figure}

At the LHC, the observation of an exclusive process of the type $pp \to p+A+p$, where a produced new heavy object $A$ is
separated from  the outgoing protons by large rapidity gaps (LRG),
 will provide very good experimental conditions to
study the properties of object $A$ \cite{KMRProsp,DKMOR,LaThuile}.
 The process is sketched in Fig.~\ref{fig:pAp}. The case of $A=H \to b\bar{b}$ is particularly interesting. Thecross is usually written in the form
\be
\sigma ~\sim~\frac{\langle S^2 \rangle}{B^2} \left|N\int \frac{dQ_t^2}{Q^4_t}f_g(x_1,x'_1,Q^2_t,\mu^2)f_g(x_2,x'_2,Q^2_t,\mu^2)\right|^2
\label{eq:d1}
\ee
where $B/2$ is the $t$-slope of the proton-Pomeron vertex, and the constant $N$ is known in terms of the $A \to gg$ decay width. The amplitude-squared factor, $|...|^2$, can be calculated in perturbative QCD, since the dominant contribution to the integral comes from the region $\Lambda^2_{QCD} \ll Q^2_t \ll M^2_A$, for the large values of $M_A^2$ of interest. The probability amplitudes, $f_g$, to find the appropriate pairs of $t$-channel gluons $(x_1,x'_1)$ and $(x_2,x'_2)$ of Fig.~\ref{fig:pAp}, are given by skewed unintegrated gluon densities at a hard scale $\mu \sim M_A/2$.
To evaluate the cross section of such an exclusive processes it is
important to know the probability, $\langle S^2 \rangle$, that the LRG survive
and will not be filled by secondaries from eikonal and enhanced
rescattering effects.
The main effect comes from the rescattering of soft partons, since they
have the largest absorptive cross sections. Therefore, we need
 a realistic model to describe soft interactions at the LHC
energy, and to predict the total cross section at LHC. The model must
account for (i) elastic rescattering (with two protons in intermediate
state), (ii) the probability of the low-mass proton excitations (with an
intermediate proton replaced by the N(1400), N(1700), etc.
resonances), and (iii) the screening corrections due to high-mass proton
dissociation.

\begin{figure} 
\begin{center}
\includegraphics[height=3.5cm]{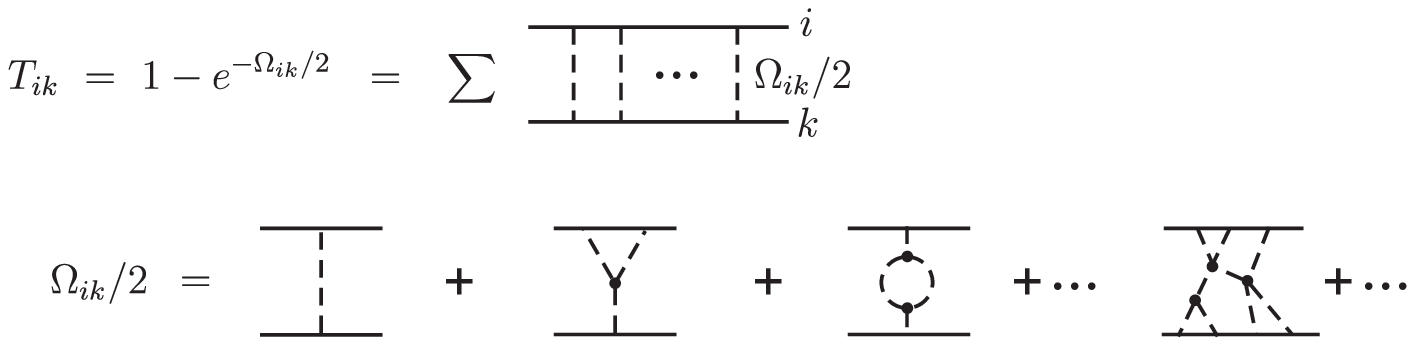}
\caption[The multi-channel]{The multi-channel eikonal form of the amplitude, where $i,k$ are diffractive (Good-Walker) eigenstates. Low-mass  proton dissociation is included by the differences
of the Pomeron couplings to one or another Good-Walker state (i) in the first
diagram, while the remaining (multi-Pomeron) diagrams on the right-hand side of the expression for $\Omega_{ik}/2$ include the high-mass dissociation.} 
\label{fig:ij}
\end{center}
\end{figure}

The effect of elastic rescattering may be evaluated in a model independent way
once the elastic $pp$-amplitude is known.
The effect of the low-mass dissociation is usually calculated in the
framework of the Good-Walker formalism\cite{GW}, that is, by introducing diffractive eigenstates, $\phi_i$ with $i=1,..n$, which only undergo `elastic' scattering. The resulting $n$-channel eikonal $\Omega_{ik}(s,b)$ depends on the energy and the impact parameter of the $pp$ interaction. The parameters of the model are chosen to reproduce the available (fixed-target and CERN-ISR) data
on the cross section of low-mass diffractive dissociation. Usually either a two- or three-channel eikonal is used. Finally, high-mass dissociation is described in terms
of Reggeon diagram technique\cite{RFT}. A symbolic representation of these soft scattering effects is shown in Fig.~\ref{fig:ij}. The latest calculations along these lines are described in Refs.~\cite{GLMM,KMRns}.
 In Ref.\cite{GLMM} the authors
account only for the triple-Pomeron vertex, and, moreover, sum up only the specific subset\footnote{For example, the third, but not the second, term on the right-hand side of the expression for $\Omega_{ik}/2$ in Fig.~\ref{fig:ij} is included; neither are multi-Pomeron terms, like the last term, included.}
of multi-Pomeron diagrams that were considered in Ref.~\cite{MPSI}, which is called the MPSI approximation. 
In Ref.~\cite{KMRns}
all possible multi-Pomeron vertices were included under a
reasonable assumption about the form of the $n \to
m$ multi-Pomeron vertices, $g^n_m$. The assumption corresponds to the
hypothesis that the screening of the $s$-channel parton $c$
during the evolution is given by the usual absorption factor
exp$(-\Omega_{ic}(b) -\Omega_{ck}(b))$, where $\Omega_{ic}(b)$ ($\Omega_{ck}(b)$)
is the value of the opacity of the beam (target) proton at impact
parameter $b$ with respect to the parton $c$.

Since the absorptive corrections increase with energy, the cross section
grows more slowly than the simple power
($\sigma\propto s^\Delta$) parametrisation \cite{DL}.
In spite of the fact that the models of \cite{GLMM} and \cite{KMRns}
are quite different to each other, after the parameters are fixed 
to describe the data on the total, elastic and single dissociation cross sections ($\sigma_{\rm tot}$, 
$d\sigma_{\rm el}/dt$ and $d\sigma_{\rm SD}/dM^2$) within the CERN-ISR -- Tevatron energy range, the latest
versions of the Tel-Aviv and Durham models
predict almost the same total cross section at the LHC, namely $\sigma_{\rm tot}\sim
90$ mb. Correspondingly, both models predict practically the same gap
survival probability $\langle S^2_{\rm eik} \rangle \sim 0.02$ with respect to the eikonal
(including the elastic and low-mass proton excitation) rescattering, for the exclusive production of a Higgs boson.

A more delicate problem is the absorptive correction to exclusive
cross sections caused by the so-called enhanced diagrams, that is by
the interaction with the intermediate partons, see Fig.~\ref{fig:pAp}. This rescattering
violates `soft-hard' factorisation, since the probability of
such an interaction depends both on the transverse momentum and on the
impact parameter of the intermediate parton.

The contribution of the first enhanced diagram was evaluated
in\cite{BBKM} in the framework of the perturbative QCD. It turns out to
be quite large.  On the other hand, such an effect is not seen
experimentally. The absorptive correction due to enhanced
screening must increase with energy. This was not observed in the
present data (see \cite{KMR-JHEP} for a more detailed discussion).

Several possible reasons are given below.

(a) We have to sum up the series of the multi-loop Pomeron diagrams.
The higher-loop contributions partly compensate the correction
caused by the first-loop graph.

(b) There should be a ``threshold'', since Pomeron vertices must be separated by a non-zero rapidity
interval \cite{KKMR}.
 That is, at present energies, the kinematical space available for the
position of a multi-Pomeron vertex in an enhanced diagram is
small, and the enhanced contribution is much less than that obtained in leading logarithmic (LL)
approximation.

(c) The factor $S^2_{\rm eik}$ already absorbs almost all the
contribution from the center of the disk. The parton only survives eikonal rescattering on the periphery, that
is at large $b$. On the other
hand, on the periphery the parton density is rather small, and the
probability of {\it enhanced} absorption is not large.
This fact can be seen in Ref.~\cite{Watt}. There, the momentum,
$Q_s$, below which we may approach saturation, was extracted from HERA data in the framework of the dipole model. Already at $b=0.6$
fm the value of $Q^2_s<0.3$ GeV$^2$ for $x<10^{-6}$.  See also
\cite{Strik1} where the value of $Q_s$ was evaluated using LO
DGLAP evolution.

Point (c) is relevant to the calculation of  $S^2_{\rm enh}$ described in \cite{GLMM}. First, note that the $b$ dependence of the beginning of `saturation caused by
enhanced graphs' is not accounted for in the MPSI approximation used
in \cite{GLMM}. In this model, we have the same two-particle irreducible amplitude (which sums up the enhanced diagrams) at any value of $b$.
Therefore, the enhanced screening effect does not depend on the
initial parton density at a particular impact parameter point $b$.
For this reason the suppression due to enhanced screening corrections
$\langle S^2_{\rm enh} \rangle=0.063$ claimed in \cite{GLMM} is much too strong\footnote{Moreover, since the irreducible amplitude approaches saturation at some fixed energy (rapidity), independent of the value
of $b$, the approximation gives
$\sigma_{\rm tot}(s\to\infty)\to$ constant. On the other hand, a theory with an
asymptotically constant cross section can only be self-consistent in
the so-called `weak coupling' regime for which the triple-Pomeron
vertex vanishes for zero momentum transfer\cite{GM}. The vertex used in
\cite{GLMM} does not vanish. This indicates that the MPSI
approximation cannot be used at asymptotically high energies, and the
region of its validity must be studied in more detail.}.

The survival factor $\langle S^2_{\rm enh} \rangle$ has also been calculated in the new version of the Durham model \cite{KMRnns}. The model includes 3 components of
the Pomeron, with the different transverse momenta $k_t$
of the partons in each Pomeron component, in order to mimic BFKL
diffusion in  $\ln~k_t$. In this way we obtain a more realistic estimate
of the `enhanced screening' in exclusive diffractive Higgs
boson production at the LHC. The model predicts 
$\langle S^2_{\rm enh}\rangle \sim 1/3$. However the CDF data on exclusive $\gamma\gamma$ and $\chi_c$ production indicate that this suppression is not so strong.

Note, that comparing the values of the survival factors in this way is too simplistic. The problem is that, with enhanced screening on intermediate partons, we no longer have exact factorisation between the hard and soft parts of the process. Thus, before computing the effect of soft absorption we must fix what is included in the bare exclusive amplitude calculated in terms of perturbative QCD.

The first observation is that the bare amplitude is calculated as a convolution of two generalised (skewed) gluon distributions with the hard subprocess matrix element, see (\ref{eq:d1}). These gluon distributions are determined from integrated gluon distributions of a global parton analysis of mainly deep inelastic scattering data. Now, the phenomenological integrated parton distributions already include the interactions of the intermediate partons with the parent proton. Thus calculations of $S_{\rm enh}$ should keep only contributions which embrace the hard matrix element of the type shown in Fig.~\ref{fig:pAp}.

The second observation is that the phenomenologically determined generalised gluon distributions, $f_g$, are usually taken at $p_t=0$ and then the observed ``total'' cross section is calculated by integrating over $p_t$ of the recoil protons assuming the an exponential behaviour $e^{-Bp_t^2}$; that is
\be
\int dp^2_t~e^{-Bp_t^2}~=~1/B~=~\langle p_t^2\rangle.
\ee
However, the total soft absorptive effect changes the $p_t$ distribution in comparison to that for the bare cross section determined from perturbative QCD. Thus the additional factor introduced by the soft interactions is not just the gap survival $S^2$, but rather the factor $S^2/B^2$ \cite{KMRxc}, which strictly speaking has the form  $S^2\langle p^2_t \rangle^2$. 

In order to compare determinations of the suppression due to absorptive effects we should compare only the values of the complete cross section for $pp \to p+A+p$. However a comparison is usually made by reducing the cross section to a factorized form. If this is done, as in (\ref{eq:d1}), then the Durham predictions for the survival factor to eikonal and enhanced screening of the exclusive production of a 120 GeV Higgs at the LHC are $\langle S^2 \rangle=0.008,~0.017, 0.030$ where enhanced sreening is only permitted outside a threshold rapidity gap $\Delta y=0,~1.5,~2.3$ respectively. The values correspond to $B=4~{\rm GeV}^{-2}$.

Let us discuss the survival factors claimed by Frankfurt et al. \cite{FS}. They use another approach.
Within the eikonal formalism, they account for elastic
rescattering only. The possibility of proton diffractive excitation
is included in terms of parton-parton correlations, for both low- and high-mass dissociation.
At a qualitative level, it is possible to consider all the effects discussed
above in terms of such a language.  On the other hand, to the
  best of our knowledge, they did not describe the available data on
$\sigma_{\rm tot},~ d\sigma_{\rm el}/dt,~ M^2d\sigma_{\rm SD}/dM^2$. Also, the energy
(i.e.  $1/x$) dependence of the parton densities was evaluated using 
simple LO DGLAP evolution. This is grossly inadequate for the low values of $x$ sampled, $x \sim 10^{-5}$.  Thus, it is difficult to
judge the accuracy of their numerical predictions.
Moreover, part of the Sudakov-like suppression, which above was calculated using perturbative QCD, is here treated as parton correlations and included in the value of $S^2_{\rm enh}$.\footnote{In general, one
may include the absence of QCD radiation in the large rapidity gap in the ``soft'' survival factors, but to make comparisons we must define precisely in which part of the calculation each effect is included.
Note also that in \cite{FS} the DL expression for Sudakov $T$--factor is used, which
grossly overestimates the suppression.} 
Therefore, one cannot compare literally the
 predictions for the gap survival factors
 $S^2=\langle S^2_{\rm eik}(b)S^2_{\rm enh}(b)\rangle$ given by \cite{FS} and by the Durham, Tel-Aviv and Petrov et al. \cite{petrov} models\footnote{The last group calculated $S^2$ within their own eikonal model and fitted the parameters in a Regge-type expression for $f_g$ to describe HERA data. The final prediction is again rather close to that by the Durham group.}. The only
 possibility is to compare the predictions for the final exclusive
 cross section. Unfortunately, such a prediction is not available in \cite{FS}.

Next, we comment on another recent calculation \cite{CDHI} along the lines of eq. (\ref{eq:d1}). They claim very large uncertainties in the predictions arising mainly from the freedom in the choice of limits of integration in the Sudakov form factor which is embedded in $f_g$. However, this is not the case. In fact, the Sudakov factors have been calculated to {\it single} log accuracy. The collinear single logarithms are summed up using the DGLAP equation. To account for the `soft' logarithms (corresponding to the emission of low energy gluons) the one-loop virtual correction to the $gg \to A$ vertex was calculated explicitly, and then the scale $\mu=0.62~M_A$ was chosen so that double log expression for the Sudakov form factor reproduces the result of the explicit calculation. Similarly, the lower limit $k^2_t=Q^2_t$ was verified to give the one-loop result. It is sufficient to calculate just the one-loop correction since it is known that the effect of `soft' gluon emission exponentiates. Thus double log expression, with $\mu=0.62~M_A$, gives the Sudakov factor to single log accuracy. Also the form used for $f_g$'s in Ref.~\cite{CDHI} contradicts the known leading log$(1/x)$ asymptotic behaviour.

Finally, we discuss a very recent calculation \cite{KL} based on the dipole approach. A new development is that instead of using a multi-channel eikonal with a fixed number of diffractive eigenstates, the authors consider an explicit wave function of a fast hadron (proton, pion) and have a continuous integration over the size of the quark-quark dipoles. In this model the incoming hadron wave function is approximated by a simple Gaussian. The parameters are fitted so as to describe the data on $\sigma_{\rm tot}, \sigma_{\rm el}$ and $F_2$ at low $x$. A shortcoming is that high-mass dissociation is calculated separately. Its contribution is not included in the proton dipole opacity $\Omega(r,b)$, for which a simplified asymptotic solution of the BFKL equation was used. Moreover, to calculate the gap survival probability, $S^2(b)$, the $b$ dependence is considered, but the dependence of the ``hard subprocess'' cross section on the dipole size
was not accounted for. That is, again, the correlation between the saturation momentum $Q_s$ and $b$ is lost. Nevertheless, the model confirms the observation that the energy dependence of $S^2$ is not too steep; $S^2$ at the LHC for central exclusive production is only reduced by a factor of about 2.5 to that at the Tevatron. Thus, Tevatron data serve as a reliable probe of the theoretical model predictions of these production rates.

In summary, we have briefly discussed various recent calculations of the exclusive process $pp \to p+A+p$ at high energy. The value of the cross section when $A=(H \to b{\bar b})$ is important for the feasibility of using tagged protons to study the Higgs sector via this process at the LHC. We have paid special attention to the survival factors of the large rapidity gaps. We see no reason to doubt the claimed value, or accuracy, of the existing predictions of the Durham model.
Recall that these predictions have been checked in many places by comparing with the available
experimental data on exclusive $\gamma\gamma$ and high $E_T$ dijet
production at the Tevatron and on exclusive diffractive $J/\psi$
production at HERA (see \cite{BH75,KMRearly} for  more details). Since all
  the factors, which enter the calculations, depend rather weakly
  (logarithmically) on the initial energy, there is no reason to
  expect that the model, which describes the data at the
  Tevatron energy, will be too far from reality at the LHC.

\end{document}